

%
%
%
%

\input amstex
\magnification=\magstep1
\headline={\ifnum\pageno=1\firstheadline\else
\ifodd\pageno\rightheadline \else\leftheadline\fi\fi}
\def\firstheadline{\hfil}
\def\rightheadline{\hfil}
\def\leftheadline{\hfil}
\footline={\ifnum\pageno=1\firstfootline\else\otherfootline\fi}
\def\firstfootline{\rm\hss\folio\hss}
\def\otherfootline{\rm\hss\folio\hss}

\font\ninebf=cmbx9
\font\ninerm=cmr9
\font\nineit=cmti9
\font\eightbf=cmbx8
\font\eightrm=cmr8
\font\eightit=cmti8

\font\eighti=cmmi8

\font\eightsy=cmsy8
\font\sixrm=cmr6
\font\sixi=cmmi6
\font\sixsy=cmsy6

\def\eightpoint{\def\rm{\fam0\eightrm}%
  \textfont0=\eightrm \scriptfont0=\sixrm
                      \scriptscriptfont0=\fiverm
  \textfont1=\eighti  \scriptfont1=\sixi
                      \scriptscriptfont1=\fivei
  \textfont2=\eightsy \scriptfont2=\sixsy
                      \scriptscriptfont2=\fivesy
  \textfont3=\tenex   \scriptfont3=\tenex
                      \scriptscriptfont3=\tenex
  \textfont\itfam=\eightit  \def\it{\fam\itfam\eightit}%
  \textfont\bffam=\eightbf  \def\bf{\fam\bffam\eightbf}%
  \normalbaselineskip=16 truept
  \setbox\strutbox=\hbox{\vrule height11pt depth5pt width0pt}}
\TagsOnRight
\baselineskip=14truept
\hsize=6.0truein
\vsize=8.5truein
\parindent=3truepc
\hfuzz=3pt


\def\e{\operatorname{e}}
\def\i{\operatorname{i}}
\chardef\ii="10
\def\bbbr{\operatorname{{I\!R}}} 
\def\bbbm{\operatorname{{I\!M}}}
\def\CD{{\Cal D}}
\def\CL{{\Cal L}}
\def\myalign{\allowdisplaybreaks\align}
\def\bFj{ {\bar F}_{(j)} }
\def\bFjp{ \bFj' }
\def\bFjpz{ \bFj^{\prime\,2} }
\def\bFjpv{ \bFj^{\prime\,4} }
\def\bFjppz{ \bFj^{\prime\prime\,2}  }
\def\ih{{\i\over\hbar}}
\def\half{{1\over2}}

\def\SU{\operatorname{SU}}

\newcount\glno
\def\plus{\advance\glno by 1}
\def\num{\the\glno}

\newcount\Refno
\def\add{\advance\Refno by 1}
\Refno=1

\edef\ABHKa{\the\Refno}\add
\edef\BJb{\the\Refno}\add
\edef\CAST{\the\Refno}\add
\edef\DEW{\the\Refno}\add
\edef\REUT{\the\Refno}\add
\edef\DURf{\the\Refno}\add
\edef\DK{\the\Refno}\add
\edef\FEYa{\the\Refno}\add
\edef\FEYb{\the\Refno}\add
\edef\FH{\the\Refno}\add
\edef\FLM{\the\Refno}\add
\edef\GENKR{\the\Refno}\add
\edef\GLJA{\the\Refno}\add
\edef\GOOb{\the\Refno}\add
\edef\GOBRb{\the\Refno}\add
\edef\GROh{\the\Refno}\add
\edef\GROj{\the\Refno}\add
\edef\GROm{\the\Refno}\add
\edef\GROr{\the\Refno}\add
\edef\GRSb{\the\Refno}\add
\edef\GRSg{\the\Refno}\add
\edef\GRSh{\the\Refno}\add
\edef\INH{\the\Refno}\add
\edef\INOWI{\the\Refno}\add
\edef\KLAU{\the\Refno}\add
\edef\KLE{\the\Refno}\add
\edef\KLEMUS{\the\Refno}\add
\edef\MORE{\the\Refno}\add
\edef\PI{\the\Refno}\add
\edef\ROEP{\the\Refno}\add
\edef\SCHU{\the\Refno}\add
\edef\SIMON{\the\Refno}\add
\edef\SISTE{\the\Refno}\add
\edef\STE{\the\Refno}\add
\edef\WIE{\the\Refno}\add

{\nopagenumbers
 \pageno=0
\centerline{DESY 92 - 189\hfill ISSN 0418-9833}
\centerline{SISSA/1/93/FM\hfill}
\centerline{December 1992\hfill}
\vskip1cm
\centerline{\ninebf CLASSIFICATION OF SOLVABLE FEYNMAN PATH INTEGRALS}
\vglue 1.0truecm
\centerline{\ninerm CHRISTIAN GROSCHE}
\centerline{\nineit Scuola Internazionale Superiore di Studi Avanzati}
\centerline{\nineit Via Beirut 4, 34014 Trieste, Italy}
\vglue 0.3truecm
\centerline{\ninerm and}
\vglue 0.3truecm
\centerline{\ninerm FRANK STEINER}
\centerline{\nineit II.Institut f\"ur Theoretische Physik,
              Universit\"at Hamburg}
\centerline{\nineit Luruper Chaussee 149, 2000 Hamburg 50, Germany}
\vglue 0.8truecm
\vfill
{\rightskip=3truepc
 \leftskip=3truepc
 \ninerm\baselineskip=12truept
 \noindent
Invited talk given at the ``Fourth International Conference on Path
Integrals from meV to MeV'', Tutzing, Germany, 1992.
To appear in the proceedings (World Scientific, Singapore).
\vglue 0.8truecm }
\vfill
\centerline{\ninerm ABSTRACT}
\vglue 0.3truecm
{\rightskip=3truepc
 \leftskip=3truepc
 \ninerm\baselineskip=12truept
 \noindent
A systematic classification of Feynman path integrals in quantum
mechanics is presented and a table of solvable path integrals is given
which reflects the progress made during the last ten years or so,
including, of course, the main contributions since the invention of the
path integral by Feynman in 1942. An outline of the general theory is
given. Explicit formul\ae\ for the so-called basic path integrals are
presented on which our general scheme to classify and calculate path
integrals in quantum mechanics is based.
\vglue 0.8truecm }

\eject}
\pageno=1

\line{\bf 1. Introduction\hfil}
\vglue 0.4truecm
During this conference (May 1992) we are celebrating the fiftieth
anniversary of the invention of path integrals in quantum mechanics,
which appear for the first time on page 35 of Feynman's thesis [\FEYa]
dated May 4, 1942. By means of his path integral [\FEYb] Feynman gave a
new formulation of quantum mechanics ``in which the central mathematical
concept is the analogue of the action in classical mechanics. It is
therefore applicable to mechanical systems whose equations of motion
cannot be put into Hamiltonian form. It is only required that some sort
of least action principle be available'' [\FEYa]. A few years later,
Feynman generalized the path integral to quantum electrodynamics and
derived from it for the first time the ``Feynman rules'' providing an
extremely effective method for performing calculations in perturbation
theory.

In this contribution we restrict ourselves to path integrals in quantum
mechanics. Until fairly recently, only a few examples of exactly
solvable path integrals were known; see the books by Feynman and Hibbs
[\FH] and by Schulman [\SCHU], which give a good account of the state
of art at the time of 1965 and 1981, respectively. However, the
situation has drastically changed during the last decade or so, and it
is no exaggeration to say that we are able to solve today  essentially
all path integrals in quantum mechanics which correspond to problems
for which the corresponding Schr\"odinger equation can be solved
exactly. (This, of course, excludes all classically chaotic systems).
It thus appears to us that the time has come to look for a systematic
classification of path integrals in quantum mechanics. A comprehensive
``Table of Feynman Path Integrals'' will appear soon [\GRSg]. In the
present short contribution we are only able to give the main idea how
our classification scheme works and which classes of path integrals are
exactly solvable. Due to lack of space, we also restrict ourselves to
purely bosonic degrees of freedom. For fermionic path integrals we have
to refer to the literature [\KLAU, \SISTE]. In the following we are
not able to give a complete list of references. A very extensive list on
the literature on path integrals comprising more than 1400 papers will
be given in our monography [\GRSh] which is in preparation.

\vglue 0.6truecm
\line{\bf 2. Formulation of the Path Integral\hfil}
\vglue 0.4truecm
Let us set up the definition of the Feynman path integral. We first
consider the simple case of a classical Lagrangian $\CL(x,\dot x)=
{m\over2}\dot x^2-V(x)$ in $D$ dimensions. Then the integral kernel
($x\in\bbbr^D$)
\plus
$$K(x'',x';t'',t')=
  \Big<x''\Big|\e^{-\i H(t''-t')/\hbar}\Big|x'\Big>\Theta(t''-t')
  \tag\num$$
of the time-evolution equation
\plus
$$\Psi(x'',t'')=\int_{\bbbr^D}K(x'',x';t'',t')\Psi(x',t')dx'\enspace,
  \tag\num$$
is represented in the form (Feynman path integral [\FEYa-\FH])
\plus
$$\myalign
  K(x'',x';t'',t')
  &=\lim_{N\to\infty}\bigg({m\over2\pi\i\epsilon\hbar}\bigg)^{ND/2}
   \prod_{j=1}^{N-1}\int_{\bbbr^D}dx_{(j)}
  \\   &\qquad\times
  \exp\left\{\ih\sum_{j=1}^N\bigg[{m\over2\epsilon}
         (x_{(j)}-x_{(j-1)})^2-\epsilon V(x_{(j)})\bigg]\right\}
  \tag\num a\\
       &
  \equiv\int\limits_{x(t')=x'}^{x(t'')=x''}\CD x(t)
  \exp\left\{\ih\int_{t'}^{t''}\bigg[{m\over2}\dot x^2
       -V(x)\bigg]dt\right\}\enspace.
  \tag\num b\endalign$$
\edef\numb{\num}%
Here we have used the abbreviations $\epsilon=(t''-t')/N\equiv T/N$,
$x_{(j)}=x(t_{(j)})$ $(t_{(j)}=t'+\epsilon j,\ j=0,\dots,N)$, and we
interpret the limit $N\to\infty$ as equivalent to $\epsilon\to0$,
$T$ fixed.

The next step is to consider a generic classical Lagrangian of the
form $\CL(q,\dot q)={m\over2}g_{ab}(q)\dot q^a\dot q^b-V(q)$ in some
$D$-dimensional Riemannian space $\bbbm$ with line element $ds^2=g_{ab}
(q)dq^adq^b$. This case, as first systematically discussed by DeWitt
[\DEW], requires a careful treatment. The Feynman path integral is most
conveniently constructed by considering the Weyl-ordering prescription
(e.g.\ [\GRSb] and references therein) in the corresponding quantum
Hamiltonian. The result then is
\plus
$$\multline
  \!\!\!\!\!\!
  K(x'',x';t'',t')=\big[g(q')g(q'')\big]^{-1/4}
  \lim_{N\to\infty}\bigg({m\over2\pi\i\epsilon\hbar}\bigg)^{ND/2}
   \prod_{j=1}^{N-1}\int_{\bbbm}dq_{(j)}
   \cdot\prod_{j=1}^N\sqrt{g(\bar q_{(j)})}
  \\   \times
  \exp\left\{\ih\sum_{j=1}^N\bigg[{m\over2\epsilon}
       g_{ab}(\bar q_{(j)})\Delta q^a_{(j)}\Delta q^b_{(j)}
       -\epsilon V(\bar q_{(j)})
       -\epsilon\Delta V(\bar q_{(j)})\bigg]\right\}
  \enspace.
  \endmultline
  \tag\num$$
\edef\numa{\num}%
Here $\bar q_{(j)}=\half(q_{(j)}+q_{(j-1)})$ denotes the midpoint
coordinate, $\Delta q_{(j)}=(q_{(j)}-q_{(j-1)})$, and $\Delta V(q)$
is a well-defined ``quantum potential'' of order $\hbar^2$ having the
form $(\Gamma_a=\partial_a\ln\sqrt{g},\ g=\det(g_{ab}))$
\plus
$$\Delta V(q)={\hbar^2\over8m}
  \Big[g^{ab}\Gamma_a\Gamma_b+2(g^{ab}\Gamma_a)_{,b}
  +{g^{ab}}_{,ab}\Big]\enspace.
  \tag\num$$
The midpoint prescription together with $\Delta V$ appears in a
completely natural way as an unavoidable consequence of the
Weyl-ordering prescription in the corresponding quantum Hamiltonian
\plus
$$\myalign
  H&=-{\hbar^2\over2m}g^{-1/2}\partial_a g^{1/2}g^{ab}\partial_b+V(q)
  \\
  &={1\over8m}\Big(g^{ab}p_ap_b+2p_ag^{ab}p_b+p_ap_bg^{ab}\Big)
     +V(q)+\Delta V(q)\enspace,
  \tag\num\endalign$$
with $p_a=-\i\hbar(\partial_a+\half\Gamma_a)$, the momentum operator
conjugate to the coordinate $q_a$ in $\bbbm$. Of course, choosing
another prescription leads to a different lattice definition in
Eq.~(\numa) and a different quantum potential $\widetilde{\Delta V}$.
However, every consistent lattice definition of Eq.~(\numa) can be
transformed into another one by carefully expanding the relevant metric
terms (integration measure- and kinetic energy term).

Indispensable tools in path integral techniques are transformation
rules. In order to avoid cumbersome notation, we restrict ourselves to
the one-di\-men\-sio\-nal case. For the general case we refer to
Refs.~[\GROm-\GRSb] and references therein. We consider the path
integral (\numb) and perform the coordinate transformation $x=F(q)$.
Implementing this transformation, one has to keep all terms of
$O(\epsilon)$ in (\numb). Expanding about midpoints, the result is
\plus
$$\multline
  \!\!\!\!\!\!
  K\big(F(q''),F(q');T\big)\!=\!\Big[F'(q'')F'(q')\Big]^{-1/2}
  \!\lim_{N\to\infty}\bigg({m\over2\pi\i\epsilon\hbar}\bigg)^{1/2}
  \prod_{j=1}^{N-1}\int dq_{(j)}\cdot\prod_{j=1}^N F'(\bar q_{(j)})
  \hfill\\   \times
  \exp\left\{\ih\sum_{j=1}^N\left[{m\over2\epsilon}
       {F'}^2(\bar q_{(j)})(\Delta q_{(j)})^2
       -\epsilon V(F(\bar q_{(j)}))-{\epsilon\hbar^2\over8m}
        {{F''}^2(\bar q_{(j)})\over{F'}^4(\bar q_{(j)})}\right]\right\}
  \enspace.
  \endmultline
  \tag\num$$
\edef\numc{\num}%
Note that the path integral (\numc) has the canonical form of the path
integral (\numa). It is not difficult to incorporate the explicitly
time-dependent coordinate transformation $x=F(q,t)$ [\GRSg, \GRSh]. Then
\plus
$$K\big(F(q'',t''),F(q',t');t'',t'\big)=A(q'',q';t'',t')
  \widetilde K(q'',q';t'',t')\enspace,
  \tag\num$$
\edef\numk{\num}%
with the prefactor
\plus
$$\multline
  A(q'',q';t'',t')=\Big[F'(q'',t'')F'(q',t')\Big]^{-1/2}
  \\   \times
  \exp\Bigg\{ {\i m\over\hbar}\bigg[
   \int^{q''}\!\! F'(z,t'')\dot F(z,t'')dz
   -\int^{q'}\!\! F'(z,t')\dot F(z,t')dz\bigg]\Bigg\}\enspace,
  \endmultline
  \tag\num$$
and the path integral representation for the kernel $\widetilde K$
given by ($\bFj=F(\bar q_{(j)},\bar t_{(j)})$,
\linebreak
$\bar t_{(j)}=\half(t_{(j)}+t_{(j-1)})$)
\plus
$$\multline
  \widetilde K(q'',q';t'',t')
  =\lim_{N\to\infty}\bigg({m\over2\pi\i\epsilon\hbar}\bigg)^{1/2}
  \prod_{j=1}^{N-1}\int dq_{(j)}\cdot\prod_{j=1}^N \bFjp
  \\   \qquad\times
  \exp\Bigg\{\ih\sum_{j=1}^N\Bigg[{m\over2\epsilon}
       \bFjpz(\Delta q_{(j)})^2-\epsilon V(\bFj)
  \hfill\\
       -{\epsilon\hbar^2\over8m}{\bFjppz\over\bFjpv}
       -\epsilon m\int^{\bar q_{(j)}}\!\! F'(z,t)\ddot F(z,t)dz
       \Bigg]\Bigg\}\enspace.
  \endmultline
  \tag\num$$
\edef\numd{\num}%
It is obvious that the path integral representation (\numd) is not
completely satisfactory. Whereas the transformed potential $V(F(q,t))$
may have a convenient form when expressed in the new coordinate $q$,
the kinetic term ${m\over2}{F'}^2\dot q^2$ is in general nasty.
Here the so-called ``time-transformation'' comes into play which leads
in combination with the ``space-transformation'' already carried out to
general ``space-time transformations'' in path integrals. The
time-transformation is implemented [\DK] by introducing a new
``pseudo-time'' $s''$ by means of
\plus
$$s''=\int_{t'}^{t''}{ds\over{F'}^2(q(s),s)}\enspace.
  \tag\num$$
A rigorous lattice derivation is far from being trivial and has been
discussed by many authors. Recent attempts to put it on a sound
footing can be found in Refs.~[\CAST, \FLM]. A convenient way to derive
the corresponding transformation formul\ae\ uses the energy dependent
Green's function $G(E)$ of the kernel $K(T)$ defined by
\plus
$$G(q'',q';E)=\bigg<q''\bigg|{\hbar\over H-E-\i\epsilon}\bigg|q'\bigg>
  =\i\int_0^\infty dT \e^{\i(E+i\epsilon)T/\hbar}
    K(q'',q';T)\enspace.
  \tag\num$$
For the path integral (\numc) one obtains the following transformation
formula (here we consider the time-independent case only)
$$\myalign
  K(x'',x';T)&=\int_{-\infty}^\infty{dE\over2\pi\i\hbar}
               \e^{-\i ET/\hbar}G(q'',q';E)\enspace,
  \tag\num\\   \global\plus
  G(q'',q';E)&=\i\Big[F'(q'')F'(q')\Big]^{1/2}
  \int_0^\infty ds''\widehat K(q'',q';s'')\enspace,
  \tag\num\endalign$$
\advance\glno by -1
\edef\numf{\num}\plus%
with the transformed path integral $\widehat K$ given by
\plus
$$\multline
  \widehat K(q'',q';s'')
  =\lim_{N\to\infty}\bigg({m\over2\pi\i\epsilon\hbar}\bigg)^{1/2}
   \prod_{j=1}^{N-1}\int dq_{(j)}
  \\   \qquad\times
  \exp\Bigg\{\ih\sum_{j=1}^N\Bigg[{m\over2\epsilon}(\Delta q_{(j)})^2
       -\epsilon {F'}^2(\bar q_{(j)})\Big(V(F(\bar q_{(j)}))-E\Big)
  \hfill\\
       -{\epsilon\hbar^2\over8m}\Bigg(
        3{{F''}^2(\bar q_{(j)})\over{F'}^2(\bar q_{(j)})}
         -2{F'''(\bar q_{(j)})\over F'(\bar q_{(j)})}
        \Bigg)\Bigg]\Bigg\}\enspace.
  \endmultline
  \tag\num$$
\edef\numg{\num}%
Further refinements are possible and general formul\ae\ of practical
interest and importance can be derived. Let us note that also an
explicitly time-dependent ``space-time transformation'' can be
formulated similarly to the formul\ae\ (\numf-\numg), c.f.\
Refs.~[\GRSg, \GRSh]. By the same technique also the separation of
variables in path integrals can be stated, c.f.\ Ref.~[\GROh].
But this will not be discussed here any further.

\vglue 0.6truecm
\line{\bf 3. Basic Path Integrals\hfil}
\vglue 0.4truecm
In this Section we present the path integrals which we consider
as the Basic Path Integrals.

\bigskip
\line{\it 3.1. Path Integral for the Harmonic Oscillator\hfil}
\smallskip
The first elementary example is the path integral for the harmonic
oscillator. It has been first evaluated by Feynman [\FEYb].
We have the identity ($x\in\bbbr$)
\plus
$$\multline
  \int\limits_{x(t')=x'}^{x(t'')=x''}\CD x(t)
  \exp\Bigg[{\i m\over2\hbar}\int_{t'}^{t''}\Big(\dot x^2
       -\omega^2x^2\Big)dt\Bigg]
  \\
  =\sqrt{m\omega\over2\pi\i\hbar\sin\omega T}
   \exp\Bigg\{{\i m\omega\over2\hbar}\bigg[({x'}^2+{x''}^2)\cot\omega T
              -{2x'x''\over\sin\omega T}\bigg]\Bigg\}\enspace.
  \endmultline
  \tag\num$$
We do not state the expansion into wave-functions ($\propto$
Hermite polynomials) which can be done by means of the Mehler formula,
nor the corresponding Green's function.
\newline
The path integral for quadratic Lagrangians can also be stated exactly
($x\in\bbbr^D$)
\plus
$$\multline
  \int\limits_{x(t')=x'}^{x(t'')=x''}\CD x(t)
  \exp\bigg(\ih\int_{t'}^{t''}\CL(x,\dot x)dt\bigg)
  \\
  =\bigg({1\over2\pi\i\hbar}\bigg)^{D/2}
  \sqrt{\det\bigg(-{\partial^2 S_{Cl}[x'',x']\over
        \partial x_a''\partial x_b'}\bigg)}
  \exp\bigg(\ih S_{Cl}[x'',x']\bigg)\enspace.
  \endmultline
  \tag\num$$
\edef\numj{\num}%
Here $\CL(x,\dot x)$ denotes any classical Lagrangian at most quadratic
in $x$ and $\dot x$, and $S_{Cl}[x'',x']=\int_{t'}^{t''}\CL(x_{Cl},
\dot x_{Cl})dt$ the corresponding classical action evaluated along the
classical solution $x_{Cl}$ satisfying the boundary conditions
$x_{Cl}(t')=x'$, $x_{Cl}(t'')=x''$. The determinant appearing in
Eq.~(\numj) is known as the van Vleck-Pauli-Morette determinant (see
e.g.\ Refs.~[\DEW, \MORE] and references therein). The explicit
evaluation of $S_{Cl}[x'',x']$ may have any degree of complexity due to
complicated classical solutions of the Euler-Lagrange equations as the
classical equations of motion.

\bigskip
\line{\it 3.2. Path Integral for the Radial Harmonic
                                           Oscillator\hfil}
\smallskip
In order to evaluate the path integral for the radial harmonic
oscillator, one has to perform a separation of the angular variables,
see Refs.~[\GOOb, \PI]. Here we are not going into the subtleties of
the Besselian functional measure due to the Bessel functions which
appear in the lattice approach [\FLM, \GOOb, \GRSb, \PI, \STE], which
is actually necessary for the explicit evaluation of the radial
harmonic oscillator path integral. One obtains (modulo the above
mentioned subtleties) ($r>0$)
\plus
$$\multline
  \int\limits_{r(t')=r'}^{r(t'')=r''}\CD r(t)
  \exp\left[\ih\int_{t'}^{t''}\bigg({m\over2}\dot r^2
  -\hbar^2{\lambda^2-{1\over4}\over2mr^2}
           -{m\over2}\omega^2r^2\bigg)dt\right]
  \\
  =\sqrt{r'r''}{m\omega\over \i\hbar\sin\omega T}
  \exp\bigg[-{m\omega\over2\i\hbar}({r'}^2+{r''}^2)\cot\omega T\bigg]
  I_\lambda\bigg({m\omega r'r''\over \i\hbar\sin\omega T}\bigg)
  \enspace,
  \endmultline
  \tag\num$$
where $I_\lambda(z)$ denotes the modified Bessel function.

\bigskip
\line{\it 3.3. Path Integral for the P\"oschl-Teller Potential
      \hfil}
\smallskip
There are two further basic path integral solutions based on the
$\SU(2)$ [\BJb, \INOWI] and $\SU(1,1)$ [\BJb] group path integration,
respectively. The first yields the path integral identity for the
solution of the P\"oschl-Teller potential according to ($0<x<\pi/2$)
\plus
$$\myalign
   \i\int_0^\infty dT&\e^{\i ET/\hbar}
  \int\limits_{x(t')=x'}^{x(t'')=x''}\CD x(t)
  \exp\left\{\ih\int_{t'}^{t''}\left[{m\over2}\dot x^2
        -{\hbar^2\over2m}\bigg({\kappa^2-{1\over4}\over\sin^2x}
   +{\lambda^2-{1\over4}\over\cos^2x}\bigg)
  \right]dt\right\}
  \\   &={m\over\hbar}\sqrt{\sin2x'\sin2x''}
  {\Gamma(m_1-L_E)\Gamma(L_E+m_1+1)\over
   \Gamma(m_1+m_2+1)\Gamma(m_1-m_2+1)}
  \\   &\qquad\times
  \bigg({1-\cos2x_<\over2}\bigg)^{(m_1-m_2)/2}
  \bigg({1+\cos2x_<\over2}\bigg)^{(m_1+m_2)/2}
  \\   &\qquad\times
  \bigg({1-\cos2x_>\over2}\bigg)^{(m_1-m_2)/2}
  \bigg({1+\cos2x_>\over2}\bigg)^{(m_1+m_2)/2}
  \\   &\qquad\times
  {_2}F_1\bigg(-L_E+m_1,L_E+m_1+1;m_1-m_2+1;{1-\cos2x_<\over2}\bigg)
  \\   &\qquad\times
  {_2}F_1\bigg(-L_E+m_1,L_E+m_1+1;m_1+m_2+1;{1+\cos2x_>\over2}\bigg)
  \tag\num\endalign$$
with $m_{1/2}=\half(\lambda\pm\kappa)$, $L_E=-\half+\half\sqrt{2mE}\,
/\hbar$, and $x_{<,>}$ the larger, smaller of $x',x''$, respectively.
${_2}F_1(a,b;c;z)$ denotes the hypergeometric function. Here we have
used the fact that it is possible to state closed expressions for the
(energy dependent) Green's functions for the P\"oschl-Teller and
modified P\"oschl-Teller potential (see below), respectively, by summing
up the spectral expansion [\KLEMUS].

\newpage
\baselineskip=13.5truept
\line{\it 3.4. Path Integral for the Modified P\"oschl-Teller
      Potential \hfil}
\smallskip
Similarly one can derive a path integral identity for the modified
P\"oschl-Teller potential. One gets ($m_{1,2}=\half(\eta\pm\sqrt{-2mE}\,
/\hbar)$, $L_\nu=\half(-1+\nu)$, $r>0$)
\plus
$$\myalign
   \i\int_0^\infty dT&\e^{\i ET/\hbar}
  \int\limits_{r(t')=r'}^{r(t'')=r''}\CD r(t)
  \exp\left\{\ih\int_{t'}^{t''}\left[{m\over2}\dot r^2
   -{\hbar^2\over2m}\bigg({\eta^2-{1\over4}\over\sinh^2r}
   -{\nu^2-{1\over4}\over\cosh^2r}\bigg)\right]dt\right\}
  \\   &={m\over\hbar}
  {\Gamma(m_1-L_\nu)\Gamma(L_\nu+m_1+1)\over
   \Gamma(m_1+m_2+1)\Gamma(m_1-m_2+1)}
  \\   &\qquad\times
  \big(\cosh r_<\big)^{-(m_1-m_2)}\big(\tanh r_<\big)^{m_1+m_2+\half}
  \\   &\qquad\times
  \big(\cosh r_>\big)^{-(m_1-m_2)}\big(\tanh r_>\big)^{m_1+m_2+\half}
  \\   &\qquad\times
  {_2}F_1\bigg(-L_\nu+m_1,L_\nu+m_1+1;m_1-m_2+1;
                          {1\over\cosh^2r_<}\bigg)
  \\   &\qquad\times
  {_2}F_1\bigg(-L_\nu+m_1,L_\nu+m_1+1;m_1+m_2+1;\tanh^2r_>\bigg)
  \enspace.
  \tag\num
  \endalign$$

\bigskip
\line{\it 3.5. General Formul\ae \hfil}
\smallskip
For the classification of solvable path integrals, one also requires
a few additional formul\ae\ which generalize the usual problems in
quantum mechanics in a specific way. Here one has e.g.
\medskip
\parindent=2truepc
\item{i)} Explicitly time-dependent problems according to e.g.\
          $V(x)\mapsto V(x/\zeta(t))/\zeta^2(t)$,
\item{ii)} Incorporation of $\delta$-function perturbation according to
          $V(x)\mapsto V(x)-\gamma\delta(x-a)$ (one dimension), and
\item{iii)} Boundary problems with impenetrable walls (half-space,
          infinite boxes) which can be derived from ii) by considering
          the limit $\gamma\to\infty$.
\medskip
\parindent=3truepc
\noindent
i) For the first class of problems, there is a general solution
provided $\zeta(t)$ has a specific form. For $\zeta(t)=(at^2+2bt+c)
          ^{1/2}$ one finds the general formula
\plus
$$\multline
   \int\limits_{x(t')=x'}^{x(t'')=x''}\CD x(t)
  \exp\left\{\ih\int_{t'}^{t''}\left[{m\over2}\dot x^2
     -{1\over\zeta^2(t)}V\bigg({x\over\zeta(t)}\bigg)\right]dt\right\}
  \\
  =\big(\zeta''\zeta'\big)^{-D/2}
  \exp\left[{\i m\over2\hbar}\left({x''}^2{\dot\zeta''\over\zeta''}
                     -{x'}^2{\dot\zeta'\over\zeta'}\right)\right]
  K_{\omega',V}\bigg({x''\over\zeta''},{x'\over\zeta'};
  \int_{t'}^{t''}{dt\over\zeta^2(t)}\bigg)\enspace,
  \endmultline
  \tag\num$$
\edef\numh{\num}%
with $\zeta'=\zeta(t')$, $\zeta''=\zeta(t'')$, etc. Here
${\omega'}^2=ac-b^2$ and $K_{\omega',V}$ denotes the path integral
\plus
$$K_{\omega',V}(z'',z';s'')
  =\int\limits_{z(0)=z'}^{z(s'')=z''}\CD z(s)
  \exp\left\{\ih\int_0^{s''}\bigg[{m\over2}\dot z^2
   -{m\over2}{\omega'}^2z^2-V(z)\bigg]ds\right\}\enspace.
  \tag\num$$

\baselineskip=14truept
Another class of time-dependent problems has a time-dependence
according to $V(x)\mapsto V(x-f(t))$. Here one gets [\DURf]
($q'=x'-f'$, $f'=f(t')$, etc.)
\plus
$$\multline
  \int\limits_{x(t')=x'}^{x(t'')=x''}\CD x(t)
  \exp\left\{\ih\int_{t'}^{t''}\bigg[
  {m\over2}\dot x^2-V(x-f(t))\bigg]dt\right\}
  \\   \qquad
  =\exp\left\{{im\over\hbar}
  \left[\dot f''(x''-f'')-\dot f'(x'-f')
  +\half\int_{t'}^{t''}\dot f^2(t)dt\right]\right\}
  \hfill\\   \qquad\qquad\times
  \int\limits_{q(t')=q'}^{q(t'')=q''}\CD q(t)
  \exp\left\{\ih\int_{t'}^{t''}\bigg[
  {m\over2}\dot q^2-V(q)-m\ddot f(t) q\bigg]dt\right\}\enspace,
  \hfill\endmultline
  \tag\num$$
\edef\numi{\num}%
Eqs.~(\numh,\numi) are special cases of Eq.~(\numk) (note that $\dot
F'(q,t)=0$ in (\numi) and therefore an additional term in the prefactor
$A(t'',t')$ appears).

\bigskip\noindent
ii) In the second class of general formul\ae\ we consider the
incorporation of $\delta$-function perturbations, i.e.\ a
$\delta$-function as an additional potential located at $x=a$ with
strength $\gamma$. However, here only a closed formula for the
corresponding Green's function can be stated; an explicit result for
the propagator can only be obtained in the simplest, or in some
exceptional cases. One obtains [\GROj]
\plus
$$\multline
   \i\int_0^\infty dT \e^{\i ET/\hbar}
  \int\limits_{x(t')=x'}^{x(t'')=x''}\CD x(t)
  \exp\left\{\ih\int_{t'}^{t''}\bigg[{m\over2}\dot x^2
       -V(x)+\gamma\delta(x-a)\bigg]dt\right\}
  \\
  =G^{(V)}(x'',x';E)+{G^{(V)}(x'',a;E)G^{(V)}(a,x';E)
      \over{\hbar\over\gamma}-G^{(V)}(a,a;E)}\enspace.
  \endmultline
  \tag\num$$
\edef\numg{\num}%
Here $G^{(V)}(E)$ denotes the Green's function for the unperturbed
problem ($\gamma=0$). Possible bound states are determined by the poles
of $G(E)$, i.e.\ by the equation $G^{(V)}(a,a,E_n)=\hbar/\gamma$.

\bigskip\noindent
iii) The third class of general formul\ae\ is obtained if we consider
in Eq.~(\numg) the limit $\gamma\to\infty$. This has the consequence
that an impenetrable wall appears at $x=a$. The result then is for the
motion in the half-space $x>a$
\plus
$$\multline
   \i\int_0^\infty dT \e^{\i ET/\hbar}
  \int\limits_{x(t')=x'}^{x(t'')=x''}\CD_{half-space}x(t)
  \exp\left\{\ih\int_{t'}^{t''}\bigg[{m\over2}\dot x^2
       -V(x)\bigg]dt\right\}
  \\
  =G^{(V)}(x'',x';E)-
   {G^{(V)}(x'',a;E)G^{(V)}(a,x';E)\over G^{(V)}(a,a;E)}
  \enspace.
  \endmultline
  \tag\num$$
Possible bound states are determined by the poles of $G(E)$, i.e.\
by the equation $G^{(V)}(a,a,E_n)=0$. Furthermore, for the motion inside
a box with boundaries at $x=a$ and $x=b$ one obtains $(a<x<b)$
\plus
$$\multline
   \i\int_0^\infty dT \e^{\i ET/\hbar}
  \int\limits_{x(t')=x'}^{x(t'')=x''}\CD_{box}x(t)
  \exp\left\{\ih\int_{t'}^{t''}\bigg[{m\over2}\dot x^2
       -V(x)\bigg]dt\right\}
  \\
  ={\left|
  \matrix G^{(V)}(x'',x';E) &G^{(V)}(x'',b;E) &G^{(V)}(x'',a;E)  \\
          G^{(V)}(b,x';E)   &G^{(V)}(b,b;E)   &G^{(V)}(b,a;E)    \\
          G^{(V)}(a,x';E)   &G^{(V)}(a,b;E)   &G^{(V)}(a,a;E)
  \endmatrix\right|\over\left|\matrix
          G^{(V)}(b,b;E)   &G^{(V)}(b,a;E)    \\
          G^{(V)}(a,b;E)   &G^{(V)}(a,a;E)\endmatrix\right|}
  \enspace.
  \endmultline
  \tag\num$$
\edef\numi{\num}%

\vglue 0.6truecm
\line{\bf 4. A Table of Exactly Solvable Feynman
                Path Integrals \hfil}
\vglue 0.4truecm
We are now in the position to present a systematic classification and a
list of exactly solvable Feynman path integrals. Of course, due to lack
of space, an actual table cannot be presented in this contribution. We
therefore list the name of the potential, respectively the name of the
quantum mechanical problem, and the basic path integrals to which the
path integrals in question can be reduced.

In our table we order the quantum mechanical problems according to
their underlying basic path integral. Of course, this classification is
closely related to the classification scheme based on Schr\"odinger's
factorization method as reviewed by Infeld and Hull [\INH],
respectively the related classification scheme of Gendenshte\u\ii n
[\GENKR] based on supersymmetric quantum mechanics.

Our classification is according to the following classes
\medskip
\parindent=2truepc
\item{i)} The general Lagrangian which is at most quadratic in $x$ and
          $\dot x$ (the harmonic oscillator being the simplest and best
          known example),
\item{ii)} The radial harmonic oscillator,
\item{iii)} The P\"oschl-Teller potential,
\item{iv)} The modified P\"oschl-Teller potential,
\item{v)} Explicitly time-dependent problems,
\item{vi)} Path integrals with $\delta$-function perturbation,
\item{vii)} Path integrals with infinite boundaries (infinite walls and
            boxes).
\medskip
\parindent=3truepc
Because of limited space, our table includes only the classes i)-iv).
A complete list will be given in [\GRSg].

\eject
\baselineskip=14truept
\hfuzz=10pt
\centerline{\bf Table of Exactly Solvable Feynman Path Integrals}
\medskip
$$\aligned
  &\vbox{\eightpoint\eightrm
         \offinterlineskip
\halign{&\vrule#&
   $\strut\ \hfil\hbox{#}\hfill\ $\cr
\noalign{\hrule}
height2pt&\omit&&\omit&&\omit&&\omit&\cr
&Quadratic Lagrangian
      &&Radial Harmonic
      &&P\"oschl-Teller
      &&Modified P\"oschl-Teller
      &\cr
&
      &&\ Oscillator
      &&\ Potential
      &&\ Potential
      &\cr
height2pt&\omit&&\omit&&\omit&&\omit&\cr
\noalign{\hrule}
\noalign{\hrule}
\noalign{\hrule}
\noalign{\hrule}
height2pt&\omit&&\omit&&\omit&&\omit&\cr
&Infinite square well
      &&Liouville mechanics
      &&Scarf potential
      &&Reflectionless potential
      &\cr
height2pt&\omit&&\omit&&\omit&&\omit&\cr
\noalign{\hrule}
height2pt&\omit&&\omit&&\omit&&\omit&\cr
&Linear potential
      &&Morse potential
      &&Symmetric top
      &&Rosen-Morse potential
      &\cr
height2pt&\omit&&\omit&&\omit&&\omit&\cr
\noalign{\hrule}
height2pt&\omit&&\omit&&\omit&&\omit&\cr
&Repelling oscillator
      &&Uniform magnetic field
      &&Magnetic top
      &&Wood-Saxon potential
      &\cr
height2pt&\omit&&\omit&&\omit&&\omit&\cr
\noalign{\hrule}
height2pt&\omit&&\omit&&\omit&&\omit&\cr
&Forced oscillator
      &&Motion in a section
      &&Spheres
      &&Hult\'en potential
      &\cr
height2pt&\omit&&\omit&&\omit&&\omit&\cr
\noalign{\hrule}
height2pt&\omit&&\omit&&\omit&&\omit&\cr
&Saddle point potential
      &&Calogero model
      &&Bispherical
      &&Manning-Rosen potential
      &\cr
&
      &&
      &&\ coordinates
      &&
      &\cr
height2pt&\omit&&\omit&&\omit&&\omit&\cr
\noalign{\hrule}
height2pt&\omit&&\omit&&\omit&&\omit&\cr
&Uniform magnetic field
      &&Aharonov-Bohm
      &&
      &&Hyperbolic Scarf potential
      &\cr
&
      &&\ problems
      &&
      &&
      &\cr
height2pt&\omit&&\omit&&\omit&&\omit&\cr
\noalign{\hrule}
height2pt&\omit&&\omit&&\omit&&\omit&\cr
&Driven coupled
      &&Coulomb potential
      &&
      &&Pseudospheres
      &\cr
&\ oscillators
      &&
      &&
      &&
      &\cr
height2pt&\omit&&\omit&&\omit&&\omit&\cr
\noalign{\hrule}
height2pt&\omit&&\omit&&\omit&&\omit&\cr
&Two-time action
      &&Coulomb-like potentials
      &&
      &&Pseudo-bispherical
      &\cr
&\ (Polaron)
      &&\ in polar and parabolic
      &&
      &&\ coordinates
      &\cr
&
      &&\ coordinates
      &&
      &&
      &\cr
height2pt&\omit&&\omit&&\omit&&\omit&\cr
\noalign{\hrule}
height2pt&\omit&&\omit&&\omit&&\omit&\cr
&Second derivative
      &&Nonrelativistic
      &&
      &&Poincar\'e disc
      &\cr
&\ Lagrangians
      &&\ monopoles
      &&
      &&
      &\cr
height2pt&\omit&&\omit&&\omit&&\omit&\cr
\noalign{\hrule}
height2pt&\omit&&\omit&&\omit&&\omit&\cr
&Semi-classical expansion
      &&Kaluza-Klein monopole
      &&
      &&Hyperbolic Strip
      &\cr
height2pt&\omit&&\omit&&\omit&&\omit&\cr
\noalign{\hrule}
height2pt&\omit&&\omit&&\omit&&\omit&\cr
&Anharmonic oscillator
      &&Poincar\'e plane
      &&
      &&Hyperbolic spaces
      &\cr
&
      &&
      &&
      &&\ of rank one
      &\cr
height2pt&\omit&&\omit&&\omit&&\omit&\cr
\noalign{\hrule}
height2pt&\omit&&\omit&&\omit&&\omit&\cr
&
      &&Hyperbolic space
      &&
      &&Kepler problem
      &\cr
&
      &&\ $+$ magnetic field
      &&
      &&\ on spheres, and on
      &\cr
&
      &&\ $+$ potentials
      &&
      &&\ pseuodspheres
      &\cr
height2pt&\omit&&\omit&&\omit&&\omit&\cr
\noalign{\hrule}
}}\endaligned$$
\hfuzz=3pt

Of course, in the case of general quantum mechanical problems, more
than just one of the basic path integral solutions is required. However,
 such problems can be conveniently put into a hierarchy according to
which of the basic path integral is the most important one for its
solution. For instance, in the path integral solution for the ring
potential (an axially symmetric Coulomb-like potential), this hierarchy
puts the radial harmonic oscillator path integral solution first,
because it requires a space-time transformation to transform the
Coulomb terms into oscillator terms.

It is obvious that all potential problems can be generalized to more
complicated problems, i.e\ one can add an additional explicit
time-dependence, implement a $\delta$-function perturbation,
respectively consider problems in half-spaces and infinite boxes, c.f.\
Eqs.~(\numh-\numi). The construction of examples is left to the reader
and can be found in [\GRSg, \GRSh].

\vglue 0.6truecm
\line{\bf 5. Summary \hfil}
\vglue 0.4truecm
In this short contribution we have sketched our approach towards a
``Table of Solvable Feynman Path Integrals''[\GRSg]. We do not claim
completeness; however, we have done our best to gather as many
information as possible. The last ten years or so have seen a lot of
activity in solving path integrals and only a few problems seem to be
left open to await an exact solution (for instance, the square well
problem, all related problems with finite discrete potential steps, and
an analysis with periodic potentials; see however Ref.~[\GOBRb], where
periodic $\delta$-functions are considered).

Since Feynman's beautiful paper [\FEYb] and his classic book written
with Hibbs [\FH], several textbooks on path integration have been
published [\ABHKa, \REUT, \GLJA, \KLE, \ROEP-\SIMON, \WIE]. Now the
time seems to be ripe for a comprehensive summary and critical review
including a systematic classification and extensive bibliography which
we are going to complete soon [\GRSg, \GRSh].

It is our hope that such a compilation of our present knowledge will
help to spread the results achieved into the physical and mathematical
community, making them available for  critical consideration and
further progress, with the ultimate goal of a comprehensive and
complete path integral description of quantum mechanics and quantum
field theory, including quantum gravity and cosmology.

\vglue 0.6truecm
\line{\bf 6. Acknowledgements \hfil}
\vglue 0.4truecm
We would like to thank the organizers of the Tutzing conference for the
nice atmosphere and warm hospitality during the conference. In
particular we want to thank C.\ DeWitt-Morette, J.\ Devreese, M.\
Gutzwiller, A.\ Inomata, J.\ Klauder, H.\ Leschke and U.\ Weiss for
interesting discussions.

\vglue 0.6truecm
\line{\bf 7. References \hfil}
\parindent=2truepc
\vglue 0.4truecm
\hfuzz=7.00pt
\medskip
\item{[\ABHKa]}
S.Albeverio, P.Blanchard and R.H\o egh-Krohn:
Some Applications of Functional Integration;
{\it Lecture Notes in Physics} {\bf 153}
({\it Sprin\-ger-Verlag}, Berlin, 1982).
\item{[\BJb]}
M.B\"ohm and G.Junker:
Path Integration Over Compact and Noncompact Rotation Groups;
{\it J.Math.Phys.}\ {\bf 28} (1987) 1978.
\item{[\CAST]}
D.P.L.Castrigiano and F.St\"ark:
New Aspects of the Path Integrational Treatment of the Coulomb
Potential;
{\it J.Math.Phys.}\ {\bf 30} (1989) 2785.
\item{[\DEW]}
B.S.DeWitt:
Dynamical Theory in Curved Spaces.~I.~A Review of the Classical and
Quantum Action Principles;
{\it Rev.Mod.Phys.}\ {\bf 29} (1957) 377.
\item{[\REUT]}
W.Dittrich and M.Reuter:
Classical and Quantum Dynamics. From Classical Paths to Path Integrals
({\it Springer-Verlag}, Berlin, 1992).
\item{[\DURf]}
I.H.Duru:
Quantum Treatment of a Class of Time-Dependent Potentials;
{\it J. Phys.A: Math.Gen.}\ {\bf 22} (1989) 4827.
\item{[\DK]}
I.H.Duru and H.Kleinert:
Solution of the Path Integral for the H-Atom;
{\it Phys. Lett.}\ {\bf B 84} (1979) 185;
Quantum Mechanics of H-Atoms from Path Integrals;
{\it Fort\-schr.Phys.}\ {\bf 30} (1982) 401.
\item{[\FEYa]}
R.P.Feynman:
The Principle of Least Action in Quantum Mechanics;
Ph.\ D.\ Thesis, Princeton University, May 1942.
\item{[\FEYb]}
R.P.Feynman:
Space-Time Approach to Non-Relativistic Quantum Mechanics;
{\it Rev.Mod.Phys.}\ {\bf 20} (1948) 367.
\item{[\FH]}
R.P.Feynman and A.Hibbs: Quantum Mechanics and Path Integrals
({\it McGraw Hill, New York}, 1965)
\item{[\FLM]}
W.Fischer, H.Leschke and P.M\"uller:
Changing Dimension and Time: Two Well-Founded and Practical Techniques
for Path Integration in Quantum Phys\-ics;
{\it J.Phys.A: Math.Gen.}\ {\bf 25} (1992) 3835.
\item{[\GENKR]}
L.\'E.Gendenshte\u\ii n:
Derivation of Exact Spectra of the Schr\"odinger Equation by Means
of Supersymmetry;
{\it JETP Lett.}\ {\bf 38} (1983) 356;
\newline
L.\'E.Gendenshte\u\ii n and I.V.Krive:
Supersymmetry in Quantum Mechanics;
{\it Sov. Phys.Usp.}\ {\bf 28} (1985) 645.
\item{[\GLJA]}
J.Glimm and A.Jaffe:
Quantum Physics: A Functional Point of View
({\it Springer-Verlag}, Berlin, 1981).
\item{[\GOOb]}
M.J.Goovaerts:
Path-Integral Evaluation of a Nonstationary Calogero Model;
{\it J.Math.Phys.}\ {\bf 16} (1975) 720.
\item{[\GOBRb]}
M.J.Goovaerts and F.Broeckx:
Analytic Treatment of a Periodic $\delta$-Function Potential in the
Path Integral Formalism;
{\it SIAM J.Appl.Math.}\ {\bf 45} (1985) 479.
\item{[\GROh]}
C.Grosche:
Separation of Variables in Path Integrals and Path Integral Solution
of Two Potentials on the Poincar\'e Upper Half-Plane;
{\it J.Phys.A: Math.Gen.}\ {\bf 23} (1990) 4885.
\item{[\GROj]}
C.Grosche:
Path Integrals for Potential Problems With $\delta$-Function
Perturbation;
{\it J.Phys.A: Math.Gen.}\ {\bf 23} (1990) 5205.
\item{[\GROm]}
C.Grosche:
Coulomb Potentials by Path Integration;
{\it Fort\-schr.Phys.}\ {\bf 40} (1992), 695.
\item{[\GROr]}
C.Grosche:
An Introduction into the Feynman Path Integral;
{\it Leipzig University preprint}, NTZ 29/92, November 1992,
unpublished.
\item{[\GRSb]}
C.Grosche and F.Steiner:
Path Integrals on Curved Manifolds;
{\it Zeitschr.Phys.}\ {\bf C 36} (1987) 699.
\item{[\GRSg]}
C.Grosche and F.Steiner:
Table of Feynman Path Integrals;
to appear in {\it Springer Tracts in Modern Physics}.
\item{[\GRSh]}
C.Grosche and F.Steiner:
Feynman Path Integrals;
to appear in {\it Springer Lecture Notes in Physics}.
\item{[\INH]}
L.Infeld and T.E.Hull:
The Factorization Method;
{\it Rev.Mod.Phys.}\ {\bf 23} (1951) 21.
\item{[\INOWI]}
A.Inomata and R.Wilson:
Path Integral Realization of a Dynamical Group;
in {\it Lecture Notes in Physics} {\bf 261}
({\it Springer-Verlag},  Berlin, 1985), p.42.
\item{[\KLAU]}
J.R.Klauder:
The Action Option and a Feynman Quantization of Spinor Fields in Terms
of Ordinary C-Numbers;
{\it Ann.Phys.(N.Y.)} {\bf 11} (1960) 123.
\item{[\KLE]}
H.Kleinert:
Path Integrals in Quantum Mechanics, Statistics and Polymer Phys\-ics
({\it World Scientific}, Singapore, 1990).
\item{[\KLEMUS]}
H.Kleinert and I.Mustapic:
Summing the Spectral Representations of P\"oschl-Teller and
Rosen-Morse Fixed-Energy Amplitudes;
{\it J.Math.Phys.}\ {\bf 33} (1992) 643.
\item{[\MORE]}
C.Morette:
On the Definition and Approximation of Feynman's Path Integrals;
{\it Phys.Rev.}\ {\bf 81} (1951) 848.
\item{[\PI]}
D.Peak and A.Inomata:
Summation Over Feynman Histories in Polar Coordinates;
{\it J.Math.Phys.}\ {\bf 10} (1969) 1422.
\item{[\ROEP]}
G.Roepstorff:
Pfadintegrale in der Quantenphysik
({\it Vieweg}, Braunschweig, 1992).
\item{[\SCHU]}
L.S.Schulman:
Techniques and Applications of Path Integration
({\it John Wiley \&\ Sons}, New York, 1981).
\item{[\SIMON]}
B.Simon:
Functional Integration and Quantum Physics
({\it Academic Press}, New York, 1979).
\item{[\SISTE]}
L.P.Singh and F.Steiner:
Fermionic Path Integrals, the Nicolai Map and the Witten Index;
{\it Phys.Lett.}\ {\bf B 166} (1986) 155.
\item{[\STE]}
F.Steiner:
Path Integrals in Polar Coordinates from eV to GeV;
in ``Bielefeld Encounters in Physics and Mathematics VII; Path
Integrals from meV to MeV'', 1985, eds.: M.C.Gutzwiller et al.~({\it
World Scientific}, Singapore, 1986), p.335.
\item{[\WIE]}
F.W.Wiegel:
Introduction to Path-Integral Methods in Physics and Polymer Science
({\it World Scientific}, Singapore, 1986).

\enddocument